\documentclass{article}
\usepackage{amsmath,amssymb,amsfonts}
\usepackage{graphicx}

\title{Estimates of the total gravitation radiation in the head-on
  black hole collision} 

\author{Osvaldo M. Moreschi \and Sergio Dain\\
Facultad de Matem\'atica Astronom\'{\i}a y F\'{\i}sica (FaMAF)\\
Universidad Nacional de C\'ordoba,\\
Ciudad Universitaria, \\
  (5000) C\'ordoba, Argentina}
\date{15 February 1996}

\begin{document}
\maketitle

\begin{abstract}
We report on calculations of the total gravitational energy radiated 
in the head-on black hole collision, where we use the geometry 
of the Robinson-Trautman metrics.

\end{abstract}

\section{Introduction}

Accurate description of the dynamics of relativistic gravitating 
systems remains one of the challenging problems in general relativity. 
There is great interest in particular on the description of systems 
involving very compact objects, as for example black holes. It 
is expected that these kind of systems will be describable in 
terms of an appropriate isolated model; which in a relativistic 
theory of gravity is expressed by an asymptotically flat spacetime. 
In this description the gravitational radiated energy will be 
encoded in the behavior of the physical field at future null 
infinity.

In numerical calculations these ideas have been used in attacking the
problem of the time symmetric asymptotically flat initial data
corresponding to a two equal mass black hole
system\cite{Smarr79}\cite{Anninos93}.  It would be nice to be able to
compare these calculations with some analytic solution of the field
equation; but in trying to do this one is faced with the fact that
there is up to now only one known vacuum asymptotically flat radiating
family of solutions, the so called Robinson-Trautman
metrics\cite{Robinson62}. These spacetimes represent the dynamics of a
single black hole which settles down, in the asymptotic
future\cite{Frittelli92}, to the Schwarzschild geometry.

In the process of two black holes falling into each other in 
a head-on collision, one can distinguish two natural eras: one 
in which the two black holes are separated and falling, and the 
other in which they already form a single black hole which is 
decaying to a stationary state. Since the Robinson-Trautman metrics 
represent a black hole, one could estimate the total gravitational 
radiation emitted after the formation of the black hole using 
these geometries. In doing this one needs to associate the data 
of the two black hole collision system to the Robinson-Trautman 
data. We do this by matching, at an arbitrary retarded time, 
the radiation content of these metrics with the radiation calculated 
from the quadruple formula with Newtonian dynamics at the moment 
of collapse.

\section{Power of the energy radiated from Newtonian dynamics + quadrupole 
formula}

Let us consider the Newtonian problem of two particles of mass $m_0$,
which start at rest at an initial distance $d_{i} $. One can estimate
the power of the energy carried away by gravitational waves when they
are at a distance d assuming Newtonian dynamics and the quadrupole
approximation. More specifically; the quadrupole radiation formula
relates the power of the energy radiated with the time derivatives of
the quadrupole:
\[
P_{Q} =\frac{1}{5} \sum\limits_{i,j=1}^{3}\left( \frac{d^{3} Q_{ij}
}{dt^{3} } \right)  ^{2}
\]
where
\[
Q_{ij} =q_{ij} -\frac{1}{3} \delta _{ij} \;q,
\]
\[
q_{ij} =\int T^{00} x^{i} x^{j}  d^{3} x,
\]
and we are using geometric units in which the Newton's gravitational 
constant and the speed of light have the unit value. Then, the 
total power radiated is
\[
P_{Q} (d)=\left( \frac{1}{15} \right) \frac{1}{\left( \frac{d}{M_{0} }
\right) ^{4} } \left( \frac{1}{\left( \frac{d}{M_{0} } \right) }
-\frac{1}{\left( \frac{d_{i} }{M_{0} } \right) } \right), 
\]
where $M_{0} \equiv 2\;m_{0} $ is the total mechanical mass and d the
distance at observation.

\section{Robinson-Trautman metrics: the quadrupole approximation}

In a previous paper\cite{Frittelli92} it was found the general
asymptotic behavior of the Robinson-Trautman metrics in the asymptotic
future. The leading behavior of these metrics is governed by the
quadrupole structure of the spacetime.

More explicitly, the whole geometry of these metrics is determined by
a scalar $V(u,\theta ,\phi )$ , where u, can be considered a retarded
time and $(\theta ,\phi )$ the coordinates of a 2-sphere. According to
Ref. \cite{Frittelli92}, for an axisymmetric quadrupole structure, the
asymptotic behavior of this scalar, for $u\rightarrow \infty $ is
\[
V=1+A\;e^{-\;\;\frac{2u}{M_{\infty } } } \;Y_{20}
+O(e^{-\;\;\frac{4u}{M_{\infty } } } ),
\]
where A is a constant $M_{\infty }$ is the asymptotic mass of the
spacetime in the regime $u\rightarrow \infty $, and $Y_{20} $ is a
spherical harmonic function. This geometry undoubtedly represents the
dynamics of a single black hole; which becomes evident when in the
limit for $u\rightarrow \infty $ , i.e., V=1, one recognizes the
Schwarzschild metric.

As we have described above, there are two natural eras in the 
process of the head-on collision: before and after the formation 
of the remaining black hole. Since the Robinson-Trautman geometries 
describe the dynamics of a single black hole, it is natural to 
try to study, with these geometries, the final stage of the process.

The time derivative of the Bondi shear of the Robinson-Trautman 
sections of future null infinity has the following asymptotic 
behavior:
\[
\dot{\sigma } _{0} =\sqrt{6} \;A\;e^{-\;\;\frac{2u}{M_{\infty } } }
\;_{2} Y_{20} +O(>),
\]
where now $_{2} Y_{20} $ is a spin weight 2 spherical harmonic
function and $O(>)$ means higher order terms of the time exponential.
With this information one can calculate the flux of gravitational
Bondi radiation at future null infinity in first order, obtaining:
\[
P_{B} =\frac{6}{4\pi } \;A^{2} \;e^{-\;\;\frac{4u}{M_{\infty } } }.
\]

To relate the Robinson-Trautman geometry with the problem of 
the head-on collision of two black holes, we identify the Bondi 
power at some retarded time with the power calculated from the 
quadrupole formula at the moment when the two black holes touch; 
that is when $d=2M_{0} $
. In this way we determine A appearing in the above 
formulae.

With the Bondi flux one can calculate in first order the total 
energy radiated, namely:
\begin{eqnarray*}
E=\int_{u_0}^{\infty}P_{B} (u)\, du &=&\left.\frac{6}{4\pi }
A^{2} \frac{M_{\infty } }{4} e^{4u/M_\infty}
 \right|_{\infty }^{u_0}\\ 
&=&\frac{6}{4\pi } A^2 \frac{M_{\infty }
}{4} e^{-4u_0 /M_\infty } =P_{B} (u_{0}
)\frac{M_\infty }{4}
.
\end{eqnarray*}
Let $M_{i} $ denote de mass of the spacetime at the \emph{initial}
retarded time; then one has the relation
\[
M_{\infty } =M_{i} -E.
\]
Therefore, the ratio radiation over initial mass, can be expressed 
by
\[
\frac{E}{M_{i} } =\frac{M_{\infty } \frac{P_{B} (u_{0} )}{4} }{M_{\infty
} \left( 1+\frac{P_{B} (u_{0} )}{4} \right) } =\frac{\frac{P_{B} (u_{0}
)}{4} }{\left( 1+\frac{P_{B} (u_{0} )}{4} \right) }
.
\]
Identifying the Bondi with the quadrupole flux one would obtain
\[
\frac{E}{M} =\frac{\left( \frac{1}{60} \right) \frac{1}{l_{f} ^{4} }
\left( \frac{1}{l_{f} } -\frac{1}{l} \right) }{1+\left( \frac{1}{60}
\right) \frac{1}{l_{f} ^{4} } \left( \frac{1}{l_{f} } -\frac{1}{l} \right)
},
\]
with $l_{f} \equiv \frac{d_{f} }{M} $
 and $l\equiv \frac{d_{i} }{M} $
. In this expression we identify M with $M_{i} $
; that is we 
map the Newtonian total mass to the total mass of the spacetime 
at the initial retarded time.

\section{Time symmetric 2-black hole initial data}

In reference \cite{Misner60} it was presented the initial data
corresponding to a time symmetric 2-black hole system. There the ADM
mass $M_{M} $ was calculated, and a magnitude $D_{M} $ (there denoted
with capital l) was defined, having the interpretation of some measure
of initial distance between the black holes. More precisely $D_{M} $
was defined as the length of the minimal line from the ``ring'' at
minimum radius from one throat to the respective ``ring'' in the other
throat.

If one wants to compare calculations coming from different realizations 
of the same physical situation, namely the head-on collision 
of two equal mass black holes, one must provide a relation between 
the physical quantities involved. The two fundamental physical 
quantities involved in this situation are the total mass of the 
system and the separation of the black holes. As we have done 
in the previous section, the identification of the masses is 
straight forward, since the natural thing to do is to identify 
the invariantly defined total masses in each case. There remains 
however the relation among the notion of distance involved in 
each case.

The question is: Is there a natural way to identify the Misner 
magnitude $D_{M} $
 with the Newtonian notion of distance appearing in 
the estimates of the total gravitational energy radiated using 
the quadrupole formula approach? A simple identification of 
 $D_{M} $ with d would be too naive in the regime of small distances. We 
seek then for some relation between 
 $D_{M} $
 and d such that it captures 
the most important physical aspects of both systems.

In order to do this let us recall that in reference \cite{Brill63} the time 
symmetric initial data for a two equal mass black hole system 
was represented by the 3-geometry:
\[
ds^{2} =\left( 1+\frac{m_{1\infty } }{2r_{1} } +\frac{m_{2\infty }
}{2r_{2} } \right) ^{4} ds_{F} ^{2}
,
\]
where $ds_{F} ^{2} $ is the line element of a flat 3-geometry, 
 $r_{1} $
 and 
 $r_{2} $
 are the 
Euclidean distances from the field point to the deleted points 
in the flat 3-geometry which represent the location of the first 
and second black hole respectively, and 
 $m_{1\infty } $
 and 
 $m_{2\infty } $
 are parameters 
which characterize the spacetime. If in the conformal, scalar 
appearing in the above line element, one eliminates the third 
term, the constant time Schwarzschild line element is obtained, 
with mass 
 $m_{1\infty } $
 expressed in terms of isotropic coordinates.

According to the arguments of reference \cite{Brill63} the mass of, let 
us say, the first black hole is
\[
m_{1} =m_{1\infty } +\frac{m_{1\infty } \;m_{2\infty } }{2r_{12} }
,
\]
where now $r_{12}$ is the Euclidean distance between the black holes; 
while the total mass of the spacetime is
\[
M=m_{1} +m_{2} -\frac{m_{1\infty } \;m_{2\infty } }{r_{12} }.
\]
This expression agrees with the first relativistic correction 
to the Newtonian total mass; since in first order one has
\[
M_{rel} \equiv M_{0} +E_{0} =m_{0} +m_{0} -\frac{m_{0} ^{2} }{d_{i} }
.
\]
Therefore it seems that the Newtonian distance, for large $d$, 
should be identified with the coordinate distance of the flat 
3-geometry of the time symmetric problem. So using the line element $ds^{2} $
 shown above for the region between the black holes, and identifying 
the radius of each black hole with double their respective masses, 
we define the distance between the black holes by
\[
D=2\;m_{1} +\int\limits_{2m_{1} }^{d-2m_{2} }\left( 1+\frac{m_{1\infty }
}{2r} +\frac{m_{2\infty } }{2(d-r)} \right) ^{2}  dr+2\;m_{2}
,
\]
which in terms of the total mass M for a two equal mass system 
can be expressed by
\[
D=2\;M\left( 1+\frac{\;M}{4d} \right) +\int\limits_{M\left(
1+\frac{\;M}{4d} \right) }^{d-M\left( 1+\frac{\;M}{4d} \right) }\left(
1+\frac{m_{\infty } }{2r} +\frac{m_{\infty } }{2(d-r)} \right) ^{2}  dr
;
\]
then, defining $L\equiv \frac{D}{M} $
 and $l\equiv \frac{d}{M} $
 one has the relation
\[
L=2\;\left( 1+\frac{\;1}{4l} \right) +\int\limits_{\left(
1+\frac{\;1}{4l} \right) }^{l-\left( 1+\frac{\;1}{4l} \right) }\left(
1+\frac{1}{4\;\lambda } +\frac{1}{4(l-\lambda )} \right) ^{2}  d\lambda
.
\]
It is observed that with respect of this variable, the black holes do
not touch exactly at $l=2$ but when $l^{*} =2\left(
  1+\frac{1}{4\;l^{*} } \right)$; which corresponds to $L=l^{*}
=2.22$. This picture is in agreement with the numerical calculations
since it is found in reference \cite{Anninos93} that for $L<2.49$ a
common apparent horizon surrounds both holes \footnote{The threshold
  mentioned in their reference is $\mu<1.36 $ which corresponds to our
  $L<2.49$.}.

We relate d to the Misner magnitude 
 $D_{M} $
 by setting 
 $D(d)=D_{M} $.

\section{Upper limit for the radiation from the area theorem}

One can estimate an upper limit for the total energy radiated 
away in the form of gravitational waves when two very separated 
black holes collide, from the area theorem; which using the relation 
between total mass and single black hole mass shown above and 
in terms of the quantities already defined, establishes that
\[
\frac{E_{rad} }{M} \leq \left( 1-\frac{1}{\sqrt{2} } \right) \left(
1+\frac{\;1}{4l} \right)
.
\]
\section{Comparison of estimates with supercomputer results}

In ref. \cite{Anninos93} the numerical calculation of the total energy
radiated for the time symmetric head-on collision of two equal mass
black holes was reported. They improved previous calculations
\cite{Smarr79} with the help of latest generation of supercomputers,
and further developed analytic and numerical techniques. In the next
graph we plot their recent\cite{Anninospriv} data, our
Robinson-Trautman (RT) estimate for the total energy radiated and the
upper limit from the area theorem.
\begin{figure}
\begin{center}
\includegraphics[width=10cm, height=6cm]{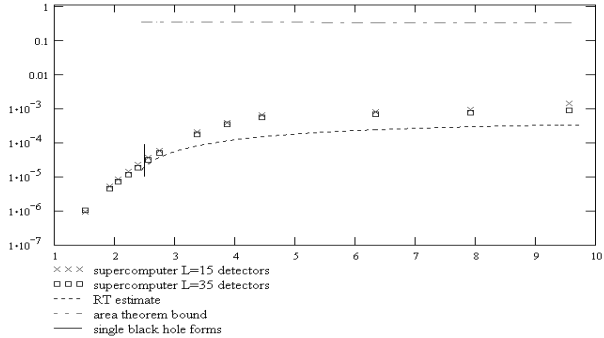}
\caption{The total gravitational energy radiated away in the process 
in units of initial total mass. The two sets of numerical data 
correspond to the location of the detectors at $L=15$ and $L=35$ 
respectively. The vertical bar at approximately $L=2.49$ corresponds 
to the region where a single black hole is formed. The RT estimate 
is plotted in terms of the relation $L(l)$ explained in the main 
text.}
\end{center}
\end{figure}

It can be observed that the RT estimate keeps bellow the supercomputer 
data; but this is no surprise since one had expected to only 
estimate, by this mean, the total energy radiated after the formation 
of the resulting black hole.

The energy radiated in the first stage, when the black holes fall from
$d_{i} $ to $d$, before the formation of the resulting black hole, can
be estimated from the quadrupole formula and using the Newtonian
dynamics; that is
\[
E_{Q} =\int\nolimits_{t_{d_{i} } }^{t_{d} }P_{Q} \left( x(t)\right)  ^{}
_{} dt =\int\nolimits_{d_{i} }^{d}P_{Q} (x)\frac{1}{\dot{x} } \;dx
;
\]
from which it is deduced that
\[
\frac{E_{Q} }{M} =\left( \frac{1}{15\sqrt{2} } \right) \left(
\frac{2}{7} \left( \frac{1}{l_{f} } -\frac{1}{l} \right)
^{\frac{7}{2} } +\frac{4}{5l} \left( \frac{1}{l_{f} }
-\frac{1}{l} \right) ^{\frac{5}{2} } +\frac{2}{3 l^{2} } \left(
\frac{1}{l_{f} } -\frac{1}{l} \right) ^{\frac{3}{2} } \right)
;
\]
with $l\equiv \frac{d_{i} }{M}$, $l_{f} \equiv \frac{d_{f} }{M}$ and M
is the total mass of the system.

Then adding to the RT estimate the quadrupole Newtonian estimate 
one obtains the total energy radiated during the entire process 
in the quadrupole approximation, which is shown in the next graph.

\begin{figure}
\begin{center}
\includegraphics[width=10cm, height=5cm]{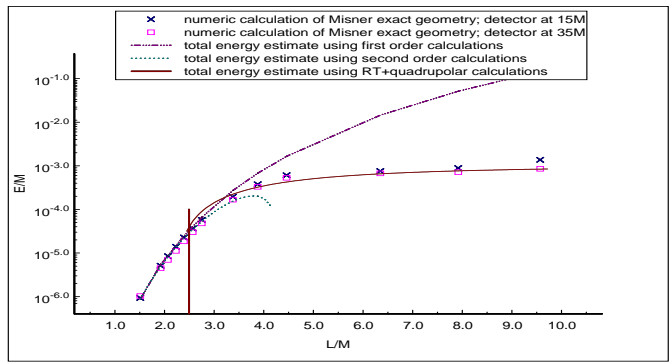}
\caption{ Same as Fig. 1 but now replacing the plain RT estimate 
by a line that shows the total gravitational energy radiated 
away in the whole process, which is the sum of the RT energy 
estimate after the formation of the remaining black hole plus 
the quadrupole Newtonian estimate of the energy radiated before 
the formation of the remanent black hole.}
\end{center}
\end{figure}

This remarkable agreement between our estimates and the supercomputer 
calculations allows one to think that the global quantities, 
as the total energy radiated, for a gravitating system like the 
one under consideration, is not very dependent on the model realization 
of the system; since, in particular, the Robinson-Trautman approach 
is not time symmetric. Therefore the supercomputer results obtained 
from the time symmetric two black holes head-on collision system 
might have a wider range of validity than the one assumed up 
to now.

It is important to note that our estimates can be calculated 
for values of L corresponding to distance over mass relation 
above the threshold of the formation of the remaining black hole; 
which from reference \cite{Anninos93} it is deduced to be around L=2.49. So 
our estimates are complementary of those calculated in reference 
\cite{Price94} in which, using a completely different approach, they have 
estimated with high precision the total energy radiated for small 
L; while their method overestimates the total energy in two orders 
of magnitude in the vicinity of 
 $L\approx 8$
(which correspond to their 
 $\mu _{0} \approx 3$
).

The total gravitational energy radiated in the large separation regime
was represented through semianalytic method in reference
\cite{Anninos93} by their equation (6). The approach followed in that
reference requires of a series of extrapolation in the applicability
of known equations. First of all they use an equation for the total
energy radiated calculated in the test particle limit\cite{Davis71}
(eq.  (1)); that is, one mass much smaller that the other mass (
$\tilde{m} <<\tilde{M} $ ) falling from infinity, and then they
extrapolate the equation to the equal mass case ( $\tilde{m}
=\tilde{M} $ ). This approach needs from several correcting factors.
Using an altered quadrupole radiation formula they suggest a
correcting factor, appearing in their eq. (4), that takes into account
the effects associated to the fact that the infall is not from
infinity but from a finite distance. The alteration in the quadrupole
formula comes from a substitution of the radial velocity by a
``nonunique'' \footnote{See reference [14] of reference
  \cite{Anninos93}.}  expression. It is observed that the integral in
their eq. (4) is calculated up to the upper limit $2\tilde{M} $ ,
which is consistent with the test particle limit approach, but it
overestimates the total gravitational energy radiation in the equal
mass case. Therefore they have to introduce another correcting factor
smaller than one, there called $F_{h}$, which is intended to take
into account the black hole structure of both objects. These points
were also studied in references \cite{Anninos95} and
\cite{Anninos95'}. It is important to emphasize that we use, without
alterations, the quadrupole radiation formula for the first stage
before the collision, up to the point in which the two horizons touch
each other ( $d\approx 4\tilde{M} $ ); this in contrast to the case of
a test particle crossing one horizon ( $d\approx 2\tilde{M} $ ). Then,
in order to take into account the black hole structure of the
remaining object we use the Robinson-Trautman geometry, which
radiation is to be added to the quadrupole estimate; this in contrast
to the reducing factor $F_{h}$.

The present estimate of the total gravitational energy radiated 
is based on a couple of simple ideas with direct physical interpretation; 
which is a contribution to the understanding of the global aspects 
of the head-on black hole collision.

It would be interesting to reproduce these calculations taken 
into account post-Newtonian techniques, as those used in reference 
\cite{Simone95}, to see whether these effects will make any substantial 
difference to our calculations.

A question that motivated us for this work was how a global quantity, 
as the total energy radiated, could be dependent on the details 
of the modeling of a physical situation. On purpose we use a 
completely different approach, and the result is that this global 
quantity does not appear to be very dependent on the details 
of the modeling.

As a final comment let us point out that since the time symmetric 
initial data obviously incorporates incoming radiation in the 
system, it is important to be able to know whether this incoming 
radiation will have an important contribution to the outgoing 
radiation. Since the Robinson-Trautman geometry can be considered 
to represent pure outgoing radiating spacetimes, and also the 
quadrupole estimate only takes into account retarded fields, 
the agreement of our estimates with the numerical calculations, 
of the time symmetric problem, suggests that the incoming radiation 
contribution to the outgoing one is negligible.

\section*{Acknowledgements}

It is a pleasure to thank Richard Price for a series of illuminating
discussions, for pointing out to us reference \cite{Brill63} and for a
detailed exposition of their work.
We are also very grateful to Peter Anninos and collaborators 
for sending us their recent calculations.


\end{document}